\documentclass{article}
\usepackage{spconf,amsmath,graphicx}

\title{Sound texture synthesis using RI spectrograms}
\name{Hugo Caracalla, Axel Roebel}
\address{UMR STMS 9912\\
Sorbonne Universit\'e, IRCAM, CNRS,\\
Paris, France}
\usepackage{booktabs}
\usepackage[table,xcdraw]{xcolor}
\usepackage{placeins}

\begin{document}
\ninept
\maketitle
\begin{abstract}
This article introduces a new parametric synthesis method for sound textures based on existing works in visual and sound texture synthesis. Starting from a base sound signal, an optimization process is performed until the cross-correlations between the feature-maps of several untrained 2D Convolutional Neural Networks (CNN) resemble those of an original sound texture. We use compressed RI spectrograms as input to the CNN: this time-frequency representation is the stacking of the real and imaginary part of the Short Time Fourier Transform (STFT) and thus implicitly contains both the magnitude and phase information, allowing for convincing syntheses of various audio events. The optimization is however performed directly on the time signal to avoid any STFT consistency issue. The results of an online perceptual evaluation are also detailed, and show that this method achieves results that are more realistic-sounding than existing parametric methods on a wide array of textures.
\end{abstract}
\section{Introduction}
\label{sec:intro}

Sound textures represent a broad class of sounds that is often overlooked despite being omnipresent in our daily lives. The hubbub of a crowd or the noise of a busy highway can all be interpreted as sound textures, acting as some relatively uniform sonic background. The question of their definition was first thoroughly investigated in \cite{saint1995classification}, and resulted in a rather strict definition that is summarized in \cite{schwarz2011} as "a superposition of small audio atoms overlapping randomly while following a higher level organization". This definition is however narrow in that it completely excludes all textures containing salient events that are not part of an overarching organization. These events might be caused by a member of a crowd coughing loudly, or of one car honking, and are more often than not present in actual texture recordings. As such we adopt a broader definition by tolerating the presence of such events as long as they are rare enough compared to the time scale of the texture organization.

Sound texture synthesis consists in creating a realistic sounding texture, and is often based on re-synthesis: given an original texture recording, the goal of the synthesis is to create a texture which style resemble that of the original, as if it had been recorded in the same conditions. Amongst existing synthesis methods, parametric methods represent a promisingly powerful paradigm. This paradigm consists in extracting from the original texture a set of parameters, and then creating a sound that possesses the same parameters. This imposition of the parameters onto a base sound is performed using an iterative optimization process. If the parametrization is done correctly, it should guarantee that the produced sound has the same textural properties as the original, without being its copy (which would defeat the purpose of the synthesis).

The synthesis method introduced in \cite{mcdermott2011sound} by McDermott \& Simoncelli is an example of such a synthesis, and uses as parameters a set of perceptual-based statistics that aim at mimicking the processing of sound textures by the human auditory system. This methods works convincingly well on a wide array of textures, but it however does not correctly reproduce salient events and impact sounds (i.e. short-lived audio events that span most of the frequency axis and have a strong attack).

Using the similarities between visual textures and the time-frequency representation of sound textures, several methods have been recently developed by adapting a successful parametric synthesis method for visual textures presented in \cite{gatys2015texture}: in this methods, the parameters are the cross-correlations between the feature map of a trained 2D Convolutional Neural Network (CNN). The method presented in \cite{ulyanov2016} by Ulyanov \& Lebedev is such an adaptation, and uses the spectrogram of a sound as input to a 1D CNN, with the frequency dimension acting as input depth, to synthesize textures with moderate success. In \cite{antognini2019audio}, this adaption is improved by Antognini \& al. with the addition of several constraints aimed at better preserving rhythmic patterns and increasing the diversity of the results. Both this and the previous method eventually synthesize a spectrogram, which is then approximately inverted using the Griffin-Lim algorithm (introduced in \cite{griffin1984signal}). Just like the method presented in \cite{mcdermott2011sound}, they also both present difficulties at synthesizing impacts. 


\section{Method}
\label{sec:method}

This difficulty encountered with impact synthesis is one of the main motivation of our work, combined with the aim of improving the overall realism of the results of parametric synthesis methods.

\subsection{Motivation}

In the state-of-the-art parametric methods introduced by McDermott \& Simoncelli in \cite{mcdermott2011sound} and by Antognini \& al. in \cite{antognini2019audio}, the presence of impacts in the original texture results in soft, watery artefacts in the synthesized texture. Our initial attempt at using a CNN-based parametrization for sound texture synthesis, presented in \cite{caracalla2019sound}, also presents similar artefacts. Because fire textures mostly contain a low rumbling sound and crackings noises, the artefacts caused by the re-synthesis of these crackings are easily perceived in the synthesized texture: the comparison between the three aforementioned parametric methods and an original fire recording is available online\footnote{See http://recherche.ircam.fr/anasyn/caracalla/icassp20/fire.php}.

Both our initial method and that of Antonini \& al. use spectrograms as input to the CNN used for parametrization. Since both are inspired by the visual texture synthesis method of \cite{gatys2015texture}, the fact that the synthesized spectrograms are visually close to the original texture spectrograms implies that the flaws of both methods originate from an ill-adapted choice of sound representation. The most obvious downside of using spectrograms for sound synthesis lies in the fact that they completely disregard the phase of the signal. This phase is recovered using the Griffin-Lim algorithm in the case of Antognini \& al., and implicitly created by performing the optimization directly onto the time signal in the case of our initial method: however, none of these methods guarantees that the phases of the different frequency bins are correlated across the spectrum. Since this correlation is most important during sharp events, managing to re-create it should thus improve impact synthesis while potentially improving the overall realism of synthesized textures. The following section presents our sound texture synthesis method which aims at rectifying this oversight.

\subsection{Analysis}

\begin{figure}[t!]
\center
\includegraphics[width=0.45\textwidth]{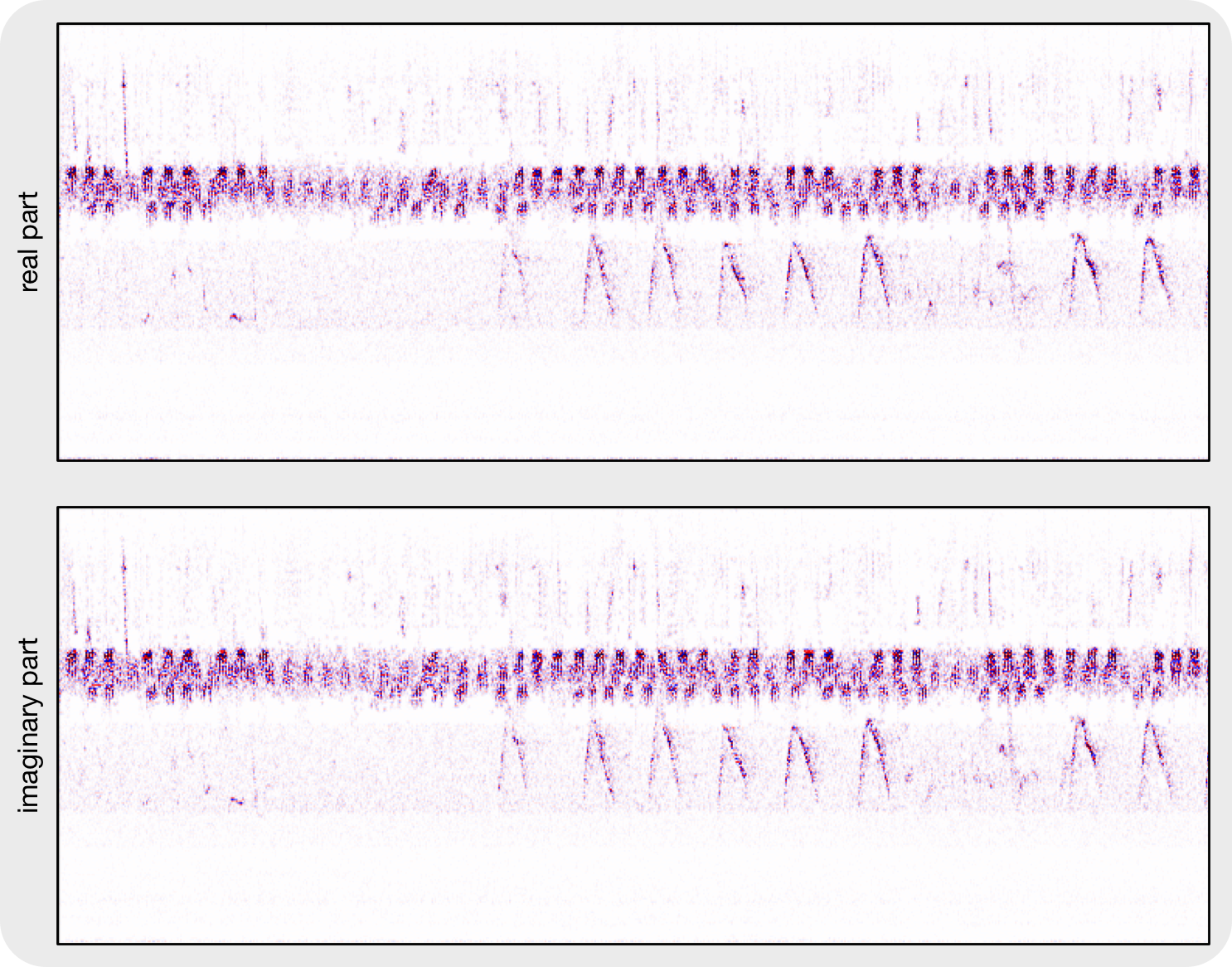}
\caption{\label{fig:reim}\textit{Visual comparison between the real and imaginary parts of a STFT using a diverging color-map.}}
\end{figure}

In this first section, we detail the process of extracting a set of parameters from a sound signal.

\subsubsection{Pre-processing}

Our aim is to use a time-frequency representation of sound that contains the phase information of the Short Time Fourier Transform (STFT) while using the paradigm of CNN-based parametric synthesis. Because phase and magnitude are strongly correlated, it is not conceivable to synthesize them separately. To bind the two, both 2D matrices can be used in the manner of color channels in the input to the CNN, as is done in \cite{engel2019gansynth}. Given how much phase matrices resemble white noise images, we however prefer using a representation in which local correlations are more visible. Instead, we thus propose the use of the real and imaginary part of the STFT, dubbed RI spectrograms in \cite{fu2017complex}. This representation, while implicitly containing both the phase and magnitude of the STFT, presents the advantage of being visibly locally correlated (as visible on Figure \ref{fig:reim}) while also resembling spectrograms enough for existing spectrogram-based texture synthesis methods to be adapted to it.

Following this reasoning, the representation that serves as input to the CNN are the compressed RI spectrograms which are organized as color channels to form a 3D matrix. All sounds worked with and presented in this article are sampled at $16$ kHz and use a window length of $512$ samples with a hop-size of $256$ for the computation of the STFT. Given the STFT $X$ of a sound signal, we first normalize it by the maximum of its absolute value and then compute the RI representation as follows:

\begin{equation}
    \begin{cases}
      R &= 2\sigma(C \operatorname{Re}(X)) - 1 \\
      I &= 2\sigma(C \operatorname{Im}(X)) - 1
    \end{cases}  
\end{equation}
with $R$ the compressed real part of the STFT, $I$ its compressed imaginary part, and $\sigma$ the sigmoid function. $C$ is a compression factor that we arbitrarily set to $10$. Defined this way, both $R$ and $I$ are always comprised between $-1$ and $1$, while being centered around $0$.

\subsubsection{Networks used}
\label{subsub:networks}

\begin{figure}[t!]
\center
\includegraphics[width=0.5\textwidth]{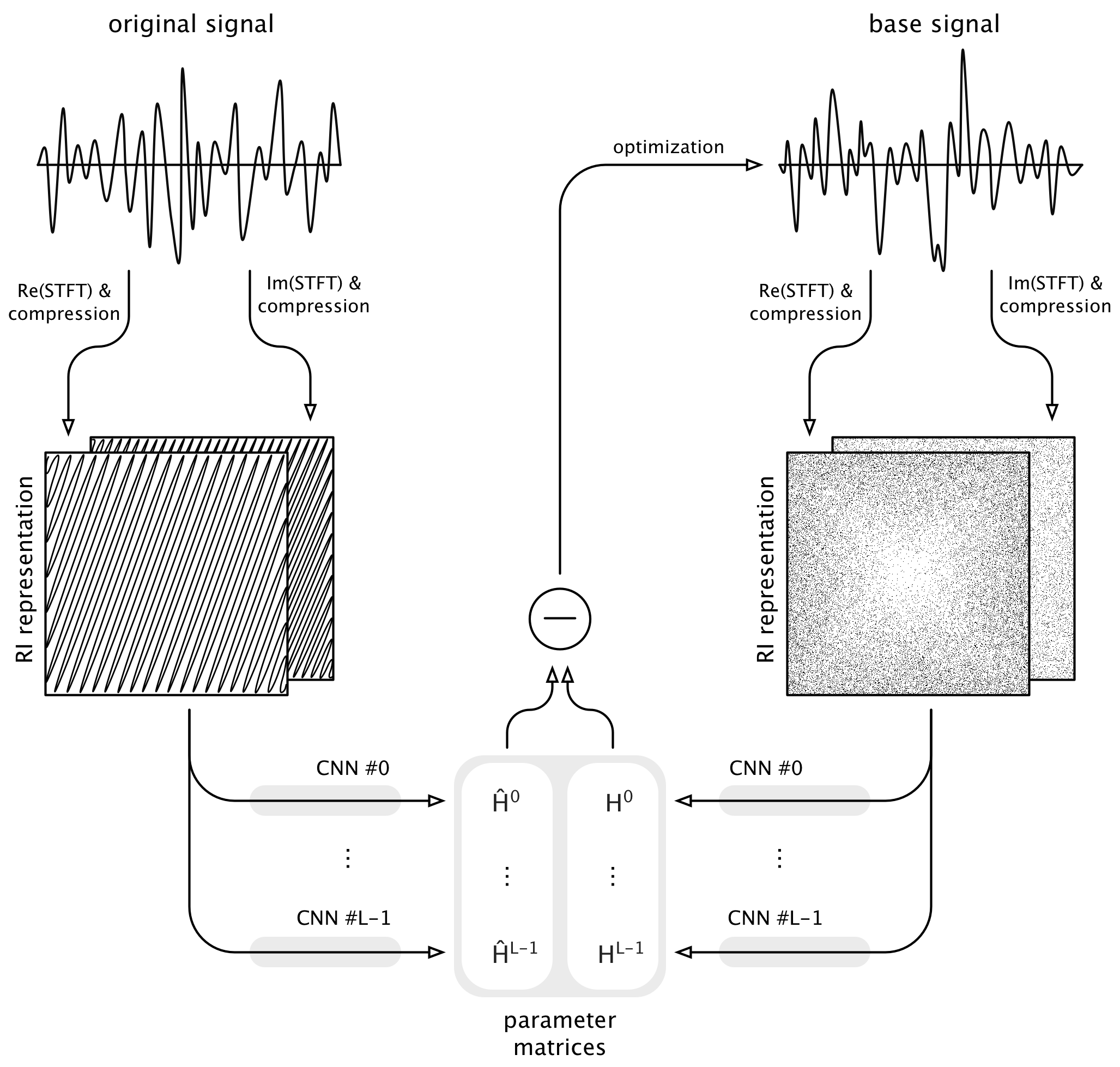}
\caption{\label{fig:cara_twin}\textit{Imposition of the statistics from the original sound texture onto a base signal. The compressed RI representation of each is computed, then passed through the same series of single-layer untrained CNNs. For both original and base signal, the cross-correlations between the feature maps of each CNN are stored inside sets of parameter matrices. The base signal is then iteratively modified until its set of parameters resemble that of the original texture.}}
\end{figure}

Similarly to Antognini \& al. in \cite{antognini2019audio}, we use 8 distinct CNN instead of one. As per \cite{ustyuzhaninov2016texture} and our own findings presented in \cite{caracalla2019sound}, given enough filters of various shapes the results obtained with trained and untrained CNN are similar in quality. Each of the 8 CNN is comprised of a single untrained convolutional layer, with $128$ filters of one unique size. These sizes are chosen following two criteria. Given that only patterns of size similar to that of the filters may be described by the parametrization, the first is that the shapes of these filters need to match events commonly present in textures. Since using larger filters means that more have to be randomly drawn to get a representative samples of possible filters, the second is that filters are better chosen as small as possible. As a consequence, we use either relatively small square filters or tall and thin filters aimed at describing impacts. The 8 filter shapes respectively used in the 8 CNN are (101, 2),  (53, 3), (11, 5), (3, 3), (5, 5), (11, 11), (19, 19), and (27, 27). All CNN also use a stride of (1, 1). No padding is applied, and all layers include a ReLU activation function. The weights of the filters are drawn from a uniform distribution between $-0.05$ and $0.05$, and no bias is applied.

\subsubsection{Parametrization}

We use the same parametrization as in \cite{caracalla2019sound}, aimed at producing a time-invariant description of textures while not being frequency-invariant. The parameters, which are the cross-correlations between the feature maps of a same CNN, are stored inside a set of parameter matrices given by:
\begin{equation}
	H^l_{ijm} = \sum_{n} F^{l}_{imn} F^{l}_{jmn}
\end{equation}
with $F^{l}_{imn}$ the feature map at position $(m, n)$ of the $i$\textsuperscript{th} filter from the $l$\textsuperscript{th} network.

\subsection{Synthesis}

Following the paradigm of parametric synthesis, a new sound possessing the same parameters as those extracted from the original texture is then created in order to obtain the new synthesized texture. This process is detailed in the following section.

\subsubsection{Texture loss}

In order to create a signal possessing the same parameters as the original texture, this process is interpreted as an optimization problem. A texture loss is defined so that it represents the distance between the parameters of a given sound and those of the original texture. A base sound is then iteratively optimized to minimize this loss. Similarly to \cite{ustyuzhaninov2016texture}, we use the following texture loss:

\begin{equation}
	\mathcal{L} = \sum_l \frac{\Vert \hat{H}^l - H^l \Vert_2}{\Vert \hat{H}^l \Vert_2}
\end{equation}
with $H^l$ the $l$\textsuperscript{th} parameter tensor with the base sound as input to the network while $\hat{H}^l$ is the $l$\textsuperscript{th} parameter tensor with the original texture as input. $\Vert . \Vert$ denotes the euclidean distance. Minimizing this loss is thus equivalent to imposing the parameters of the original texture onto the base sound.

\subsubsection{Parameters imposition}

Given a synthesized RI spectrogram, it is possible to invert the corresponding STFT and produce a time signal with no need for a phase retrieval process. However, it is important to keep in mind that the complex matrix obtained by recombining the real and imaginary parts has no guarantee of being consistent\footnote{Consistency in the sense of the STFT is the property of being the image by the STFT of a 1D signal. Not all complex matrix are consistent: inverting them is still possible, but the STFT of the resulting signal will not be identical to the initial complex matrix.}. In order to avoid this issue, and using a method that can also be found in \cite{caracalla2019sound, barry2018style, tomczakaudio}, we instead directly modify the base signal in the time domain. The resulting method is illustrated in Figure \ref{fig:cara_twin}.

In practice, we use the L-BFGS optimization algorithm and the tensorflow library to perform the iterative imposition of the parameters. In order to synthesize textures of 7 seconds, the optimization is carried out in 5000 steps and lasts roughly 7 minutes on a GeForce GTX 1080 Ti GPU.

\subsection{Results}
\label{sub:meth_res}

Examples of sounds synthesized using an array of various original are available online\footnote{See http://recherche.ircam.fr/anasyn/caracalla/icassp20/results.php}. As is clearly audible on the "fire" texture, our method succeeds in convincingly re-synthesizing impacts due to the combination of the use of tall filters and the RI representation. In addition to this, it manages to synthesize noisy textures (such as the "bees" and "static" texture), pitched events (such as the "birds" texture) and salient, un-recurring events (such as the cutlery noises in "crowd") in a very convincing way.

However, the "wind" texture shows a limit of this method: due to the size of the filters of the CNN described in Section \ref{subsub:networks}, the characteristic size of the events it may reproduce is approximately of 0.5 seconds. The texture "wind", however, is the only one that contains an event (in this case, the howling), which characteristic time is of several seconds. As a result, the texture synthesized by our method does not manage to reproduce the slow evolution of the howling. This behavior could be changed by horizontally extending the filters of its CNN, although doing this would mean reproducing longer patches of the original: we would thus risk creating a synthesized sound resembling the original texture too much.

\section{Perceptual evaluation}

\label{sec:eval}

\begin{table*}[t!]
\centering
\begin{tabular}{@{}lllllllllll@{}}
\toprule
 & \multicolumn{1}{c}{applause} & \multicolumn{1}{c}{bees} & \multicolumn{1}{c}{birds} & \multicolumn{1}{c}{crowd} & \multicolumn{1}{c}{fire} & \multicolumn{1}{c}{insects} & \multicolumn{1}{c}{rain} & \multicolumn{1}{c}{sink} & \multicolumn{1}{c}{static} & \multicolumn{1}{c}{wind} \\ \midrule
hidden & \cellcolor[HTML]{EFEFEF}1.64 & \cellcolor[HTML]{EFEFEF}2.56 & \cellcolor[HTML]{EFEFEF}2.40 & \cellcolor[HTML]{EFEFEF}1.80 & \cellcolor[HTML]{EFEFEF}2.01 & \cellcolor[HTML]{EFEFEF}3.55 & \cellcolor[HTML]{EFEFEF}1.99 & \cellcolor[HTML]{EFEFEF}2.55 & \cellcolor[HTML]{EFEFEF}2.55 & \cellcolor[HTML]{EFEFEF}1.56 \\
RI & \cellcolor[HTML]{D6D6D6}\textbf{2.19} & \cellcolor[HTML]{D6D6D6}\textbf{2.40} & \cellcolor[HTML]{D6D6D6}2.94 & \cellcolor[HTML]{D6D6D6}\textbf{2.20} & \cellcolor[HTML]{D6D6D6}\textbf{2.10} & \cellcolor[HTML]{D6D6D6}\textbf{2.74} & \cellcolor[HTML]{D6D6D6}\textbf{2.18} & \cellcolor[HTML]{D6D6D6}\textbf{2.24} & \cellcolor[HTML]{D6D6D6}\textbf{2.46} & \cellcolor[HTML]{D6D6D6}4.75 \\
spec & 5.93 & 6.96 & 4.21 & 5.46 & 5.05 & 5.40 & 6.58 & 5.79 & 5.18 & 5.11 \\
antognini & 4.74 & 4.07 & \textbf{2.76} & 2.33 & 5.94 & 4.94 & 5.07 & 5.67 & 5.15 & 5.26 \\
ulyanov & 4.77 & 3.65 & 4.39 & 5.96 & 6.38 & 3.57 & 3.71 & 5.19 & 4.44 & 6.81 \\
mcdermott & 4.31 & 4.48 & 5.43 & 5.52 & 3.56 & 4.70 & 5.03 & 3.14 & 5.38 & \textbf{2.73} \\
anchor & \cellcolor[HTML]{EFEFEF}7.93 & \cellcolor[HTML]{EFEFEF}6.35 & \cellcolor[HTML]{EFEFEF}7.94 & \cellcolor[HTML]{EFEFEF}7.75 & \cellcolor[HTML]{EFEFEF}7.65 & \cellcolor[HTML]{EFEFEF}7.83 & \cellcolor[HTML]{EFEFEF}7.74 & \cellcolor[HTML]{EFEFEF}7.83 & \cellcolor[HTML]{EFEFEF}7.43 & \cellcolor[HTML]{EFEFEF}4.21 \\ \bottomrule
\end{tabular}
\caption{\label{tab:glob}\textit{Mean rankings of all methods across all textures: the rankings range from 1 (preferred) to 8 (rejected). For each texture, the best-ranked method outside of the hidden reference is denoted by a bold score.}}
\end{table*}

\begin{figure}[t!]
\center
\includegraphics[width=0.42\textwidth]{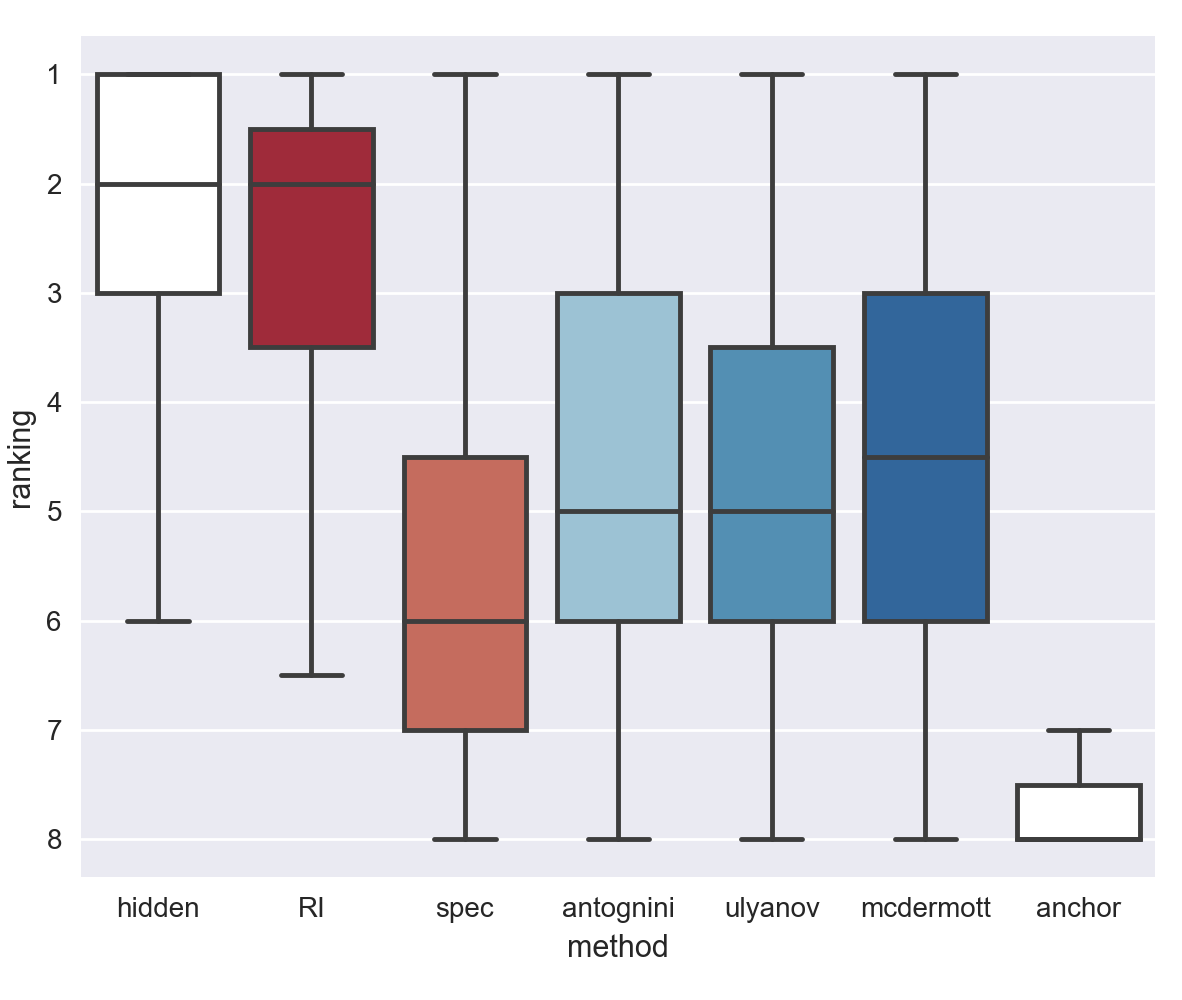}
\caption{\label{fig:glob_rank}\textit{Rankings of the different methods across all textures: hidden reference and anchors are colored in white, our methods in shades of red and state of the art methods in shades of blue.}}
\end{figure}

In order to assess the realism of textures synthesized using our method, we performed an online perceptual evaluation comparing it to other state-of-the-art methods.

\subsection{Experimental set-up}

Our protocol was loosely based on MUSHRA (Multiple Stimuli with Hidden Reference and Anchor), defined in \cite{mushra}, in which an original sample is compared to several test samples. A hidden reference (a "perfect" sample) and an anchor (a "bad" sample) are also hidden among the test samples in order to act as references. Although the original texture might have been used as the hidden reference (referred to as \textit{hidden} from now on), doing so would have blurred the distinction between identity and similarity: instead we decided to use its continuation as hidden reference. We created each anchor (\textit{anchor}) by filtering a white noise so that it had the same frequency spectrum as its corresponding original texture.

With the kind consent of Dr. Joseph M. Antognini, we used his sound set available online\footnote{See https://antognini-google.github.io/audio\_textures/baselines.html} to choose our original samples from. In addition to containing a wide array of textures, this set also contained the synthesized versions of each texture using the methods presented in \cite{antognini2019audio} (\textit{antognini}), \cite{ulyanov2016} (\textit{ulyanov}) and \cite{mcdermott2011sound} (\textit{mcdermott}). From this set we chose 10 original textures: this selection was made so as to cover an array of textures as broad as possible. We re-synthesized those textures using our RI-based method (\textit{RI}), but also our previous spectrogram-based method presented in \cite{caracalla2019sound} (\textit{spec}) for comparison's sake.

All sounds were down-sampled when needed so that they all had a sample rate of 16 kHz. They were also cropped to a length of 4 seconds. From the 7-second long selected textures, the first 4 seconds were used as original textures for the test and the last 4 seconds as hidden references. Because uneven audio volumes may influence the perception of artefacts, we normalized all samples so that their energy (or variance) were identical. All sounds can be listened to online\footnote{See http://recherche.ircam.fr/anasyn/caracalla/icassp20/assets.php}.

For each of the selected textures, the participants were presented with the original sample followed by the 7 corresponding test samples (\textit{hidden} and \textit{anchor}, \textit{antognini}, \textit{ulyanov}, \textit{mcdermott}, \textit{RI} and \textit{spec}). Like in MUSHRA, they were asked to rate how similar sounding each texture was to the original on a scale ranging from 0 (unrecognizable) to 100 (perfect): it was stressed in the instructions that the goal of these synthesis methods was not to reproduce an identical copy but a sample appearing to have been recorded moments later. The name and order of both textures and methods were anonymized so as to prevent the introduction of any bias.


\subsection{Results}

A total of 64 valid and full evaluations were filled at the time of the writing of this article. Due to the various rating strategies adopted by participants, we find that the rankings of the different methods for each texture are more telling and more stable than grades: as such, we use those as metric. The rankings range from 1 (preferred) to 8 (rejected), and an average ranking is given when several method have the same grade.

The mean rankings for each texture are displayed in Table \ref{tab:glob}, while the global rankings of the different method on all textures are shown on Figure \ref{fig:glob_rank}. We use box plots as a way to display data without making any assumption regarding its statistical distribution.

The high rankings of the hidden reference, shown by its high mean across all texture, are encouraging as they show that participants correctly understood the task given to them. The rankings of \textit{RI} being close to those of the hidden reference is an extremely positive result regarding the realism of our method. The difference between these rankings and those of \textit{spec} is also a concrete proof of the improvements that the changes in time-frequency representation and CNN architecture bring. Hidden reference put aside, our \textit{RI} methods ranks first on all textures outside of "birds", for which it ranks slightly behind \textit{antognini}, and "wind", for which it ranks behind \textit{mcdermott} and \textit{anchor}. This behavior was expected given the discussion of our results presented in Section \ref{sub:meth_res}, although the fact that a simple filtered white noise is ranked this high compared to state-of-the-art methods was rather unexpected.

Those results confirm that in addition to succeeding in the synthesis of impacts (present in "fire" and "applause"), our methods also manages to synthesize monotonous textures (such as "bees" or "static") and textures that present salient events (such as "crowd") with a realism that is comparable to that of an actual recording and surpasses current state-of-the-art methods in parametric sound texture synthesis.

\section{Conclusion}

We have demonstrated that the use of a more fitting time-frequency representation, coupled with a careful choice of filter shapes, allows for a more convincing CNN-based parametric sound texture synthesis. This improvement has been further assessed by an online perceptual evaluation which compared original texture samples with samples synthesized using both our algorithm and several state-of-the-art parametric synthesis methods: its results have been unequivocally in favor of our algorithm, showing that our samples were rated similarly to original texture samples. Despite giving less convincing results on a slowly evolving texture, we are confident that the architecture of the CNN used in our method can be investigated further to also work with this kind of texture.

Overall, this shows that our parametrization is suited to the description and synthesis of sound textures. From there, further investigations might be performed so as to test the influence of the manipulation of these parameters on the audio signal: attempts at texture control or at (textural) style transfer could for instance be made.


\bibliographystyle{IEEEbib}
\bibliography{strings,refs}

\end{document}